\documentclass[smallextended]{svjour3}       
\smartqed  
\usepackage{graphicx,natbib,amsmath,xcolor,hyperref}

\usepackage{multirow}
\newcommand{\diff}{\text{\rm d}}

\newcommand{\Bxi}    {\ensuremath{\boldsymbol\xi}}
\newcommand{\Beta}   {\ensuremath{\boldsymbol\eta}}

\newcommand{\Bb}{{\boldsymbol{\mathnormal b}}}

\newcommand{\Be}{{\boldsymbol{\mathnormal e}}}
\newcommand{\Bf}{{\boldsymbol{\mathnormal f}}}

\newcommand{\Bu}{{\boldsymbol{\mathnormal u}}}

\newcommand{\Bx}{{\boldsymbol{\mathnormal x}}}
\newcommand{\By}{{\boldsymbol{\mathnormal y}}}

\newcommand \MZ [1] {\bgroup\noindent[\textcolor{blue}{\textbf{MZ}: #1}]\egroup\ignorespacesafterend}
\newcommand \SG [1] {\bgroup\noindent[\textcolor{violet}{\textbf{SG}: #1}]\egroup\ignorespacesafterend}

\begin{document}

\title{An intermediately-homogenized peridynamics approach to failure of microstructually disordered materials}

\titlerunning{Peridynamic modelling of failure in disordered materials}        

\author{Shucheta Shegufta         \and
        Michael Zaiser 
}


\institute{S. Shegufta and M. Zaiser\at
           Friedrich-Alexander Universit\"at Erlangen-N\"urnberg \\
           Department of Materials Science, WW8-Materials Simulation\\
           Dr.-Mack Strasse 77, 90762 F\"urth, Germany
           \email{shucheta.shegufta@fau.de}           
}

\date{Received: date / Accepted: date}

\maketitle

\begin{abstract}
Peridynamics provides a versatile tool for fracture modelling in materials where fracture pathways cannot be predicted beforehand, but must be envisaged as an emergent features of the deformation process. One class of materials where this is surely the case are materials with strong microstructural disorder such as random composites, random porous materials or disordered metamaterials. For this class of materials we propose an intermediately-homogenized peridynamic modelling approach where the disordered microstructure is not resolved in full spatial detail but described in terms of random order parameter fields which retain essential information about the local heterogeneity and spatial correlations of material properties. 

\end{abstract}

\maketitle

\section{Introduction}

The microstructure of most real materials is characterized by randomness and disorder. Such randomness may arise on multiple scales: on the level of the arrangement of atoms in alloys, we may observe amorphous structures without any underlying long range order (for overview see e.g. \citet{qiao2019structural}) or chemically disordered alloys where atoms of different chemical type are arranged randomly on a crystal lattice (see e.g. \citet{li2021mechanical}). On larger scales, disorder may be manifest in the arrangement of defects such as dislocations or point defects \citep{wu2025dislocations}, and in the grain microstructure of polycrystals whose statistical modelling has been an important driver for the development of homogenization and effective medium methods in mechanics of materials \citep{kroner1986statistical}. Finally, we observe random geometrical heterogeneities on the macroscale, e.g. in terms of randomly distributed pores or inclusions \citep{su2023statistical,ritter2023effects}. The presence of disorder has often strong effects on the deformation and failure behavior of materials, as weak spots may facilitate microcrack nucleation and damage accumulation, whereas strong regions of the microstructure may promote crack arrest. To exploit these features, disorder may even be introduced 'by design' to engineer the failure modes of architected metamaterials \citep{zaiser2023disordered} and enhance their mechanical resilience. 

Peridynamics is a modelling approach towards materials deformation and fracture which is particularly suitable for complex crack propagation problems where the crack path cannot be easily predicted beforehand, and where a unified treatment of crack nucleation and propagation is required. The emergence of non-trivial, complex fracture surfaces and fracture pathways characterized by multiple instances of crack nucleation, crack branching, and crack arrest are, at the same time, generic features of failure processes in disordered materials, for overview articles see e.g. \citet{alava2006statistical} or \citet{herrmann2014statistical}. This makes peridynamics a natural approach towards modelling fracture and failure of materials with disordered microstructures. 

Peridynamics modelling of disordered materials has followed two main lines: {\em Fully resolved} models that aim at fully capturing the geometrical features of the disordered microstructure, and {\em fully homogenized} models that aim at replacing the random heterogeneous microstructure by a homogeneous effective medium. 

Fully homogenized models build upon the powerful framework of homogenization theory in classical mechanics and extend this to the structure of a peridynamic theory (see e.g. \citet{buryachenko2017effective,buryachenko2020generalized,buryachenko2023effective}). However, such homogenized models fail to capture the effects of microstructural heterogeneity on crack propagation and can therefore not be used in situations where variability of crack paths on the scale of the microstructure must be considered. Moreover homogenization approaches, while well developed 
for elasticity and other linear properties, are not well adapted to problems such as fracture which may be governed by statistics of extremes. 

On the other side, fully resolved models describe the random microstructure, on the considered scale, in full geometrical detail. There are very many studies of this type and the approach has been applied to all kinds of random microstructures, from porous media \citep{ritter2023effects} over concrete \citep{shi2021peridynamics,geng2025peridynamics} to multiphase composites \cite{nayak2020peridynamics}. The main problem of this approach is that full spatial resolution may include a lot of detail which is irrelevant to the actual response but increases the computational cost. This observation leads to the idea of developing intermediately homogenized models, which  consider stochastic heterogeneity of the microstructure, but describe this not in terms of actual, fully geometrically resolved microstructures but in terms of random fields that reflect statistical properties of the actual microstructure. In principle such models can be obtained by coarse graining a model with fine resolution, and determine the parameters of the coarse grained model from simulations of the finer scale model, see \citet{silling2024peridynamic}. Such parameterizations can be remarkably faithful in reproducing, for individual realizations of the disordered microstructure, the behavior of the fine scaled model on the coarser scale \citep{silling2024peridynamic}, However, if one wants to screen over different realizations of the disordered microstructure to establish statistical signatures of the failure process (e.g., a statistical size effect), then each realization requires a different parameterization. This renders any advantage of coarse graining in view of computational efficiency tenuous. 

To remedy this problem, one may consider models which treat randomness and heterogeneity in a phenomenological manner, by describing the microstructure in terms of stochastic variables whose statistics is characterized by a limited number of parameters, to be fitted {\em a posteriori} to experimental data or lower-scale model data. In the context of Peridynamics, such a model has been formulated for porous media by \citet{chen2019peridynamic}, who mimic porosity (a statistical signature) by random deletion of a given fraction of bonds from a bond-based peridynamic model. Since the number of bonds connecting to any given material point then becomes a random number, this process leads to random fluctuations in local modulus and local strength. Variants of this model, called an {\em intermediately homogenized} model by its authors, have been applied to a range of problems including fracture of concrete \citep{wu2021stochastically} and fiber-reinforced composites \citep{mehrmashhadi2019stochastically}, and wave propagation and other mechanical properties of porous solids \citep{ozdemir2023evaluation}. However, it was pointed out by \citet{shegufta2025peridynamics} that the description of disorder by \citet{chen2019peridynamic} suffers from a severe conceptual problem. If we independently remove a given fraction of the bonds to introduce disorder, then local fluctuations in elastic modulus and failure strength arise from the fluctuations in the number of bonds connecting to any given material point. However, it is easy to see that, in the continuum limit where the number of bonds connecting to any material point is infinite, the relative fluctuations of this number go to zero when bonds are deleted independently. Thus, the intermediately homogenizted model of  \citet{chen2019peridynamic} is not $m$ convergent as, in the continuum limit, it converges to a disorder-free, fully homogenized material. As a consequence, any modelling of disorder effects in this model is conditional on the choice of a numerical discretization parameter, such as the number of collocation points within the peridynamic horizon. 

In the present study, we propose an alternative approach to intermediately-homogenized peridynamic modelling of microstructural disorder. Our approach is based on the idea of considering mechanical properties as random fields which describe the microstructure in terms of a spatial stochastic process with prescribed probabilities and spatial correlations. We present the model in Section 2 and demonstrate how the model of \citet{chen2019peridynamic} can be recovered as a limiting case. We use the model to perform simulations (we use snow as a test case) and discuss numerical aspects ($m$ convergence, convergence to the classical limit of vanishing horizon) as well as physical aspects concerning the dependency of fracture properties on the degree of disorder, disorder correlations, and sample size. We conclude with an outlook regarding possible generalizations of the model. 

\section{Intermediately homogenized model: Formulation}

\subsection{Basic model structure}

We consider a standard bond-based peridynamics formalism for a brittle material as originally defined by \citet{silling2000reformulation}, see also \citet{silling2005meshfree}. Our presentation follows the notation of \citet{bobaru2016handbook}. The displacement field is defined by the $d$-dimensional vector $\Bu(\Bx) = \By(\Bx) - \Bx$ where $\By(\Bx)$ is the current location of the point with material coordinates $\Bx$.  The force balance equation for the point $\Bx$ is given by
\begin{equation}
	\rho(\Bx) \ddot{\Bu}(\Bx) = \int_{{\cal H}_{\Bx}} \Bf \left(\Bx,\Bx^{\prime} \right) \diff \Bx^{\prime} + \Bb(\Bx) ,
    \label{eq:forcebalance}
\end{equation}
where $\Bf \left( \Bx,\Bx^{\prime} \right)$ is the pair force between $\Bx^{\prime}$ and $\Bx$, $\Bb$ is a body force field, and interactions are restricted to a family ${\cal H}_{\Bx}$ which we take to be a circle of radius $\delta$, the so-called horizon, around $\Bx$, $\left\vert \Bx - \Bx^* \right\vert \le \delta \; \forall \; \Bx^* \in {\cal H}_{\Bx}$. 

For specifying the pair force, we introduce the notations $\Bxi \left(\Bx,\Bx^{\prime} \right) = \Bx^{\prime} - \Bx$ and $\Beta \left(\Bx,\Bx^{\prime} \right) = \Bu \left(\Bx^{\prime} \right) - \Bu \left(\Bx \right)$. The pair force is then taken linearly proportional to the bond stretch $s$, and pointing in the direction of the vector $\Be_u$ connecting both points in the current configuration:
\begin{equation}
	\Bf \left(\Bx,\Bx^{\prime} \right)= c \left(\Bxi \right) s \Be_u \left(\Bxi,\Beta \right) \quad,\quad s = \frac{\left\vert \Beta + \Bxi \right\vert - \left\vert \Bxi \right\vert}{|\Bxi|}\quad,\quad \Be_u \left(\Bxi,\Beta \right) = \frac{\Beta + \Bxi}{ \left\vert \Beta+\Bxi \right\vert }
\end{equation}
Here $c(\Bxi)$ is the so-called bond micro-modulus which for a homogeneous and isotropic bulk material depends on the bond length $\xi = |\Bxi|$ only. In the following we take such a material as a reference medium and assume for this case a constant micro-modulus, $c(\xi)=c_0$, which we choose such that the behavior of the material under homogeneous small deformations matches an isotropic linear-elastic material of bulk modulus $k_d$ in $d$ dimensions, hence 
\begin{equation}
    c_0 = \left\{ \begin{array}{ll}
    \frac{12k_d}{\pi\delta^3}, \quad d=2,\\[6pt]
    \frac{18k_d}{\pi\delta^4}, \quad d=3.
    \end {array}\right.
    \label{eq:bondModulus}
\end{equation}
where $d$ is the spatial dimensionality \citep{bobaru2016handbook}. The bond energy can then be written in terms of the bond length $\xi$ and bond stretch $s$ as
\begin{equation}
	e \left(\Bx,\Bx^{\prime}\right) = 	e\left(\Bx^{\prime},\Bx\right) = \frac{c_0}{2} s^2 \xi.
\end{equation}
Elastic-brittle behavior is introduced by defining a critical bond stretch $s_{\rm c}$ in such a manner that, once $s > s_{\rm c}$, the bond fails: in this case, its micro-modulus is irreversibly set to zero. The elastic energy $\frac{c_0}{2} s_{\rm c}^2 \xi$ of the failed bond is then converted into defect energy $E_{\rm d}$; for our considerations it is irrelevant whether the defect energy is considered as some kind of microscopically stored, non recoverable internal energy, or simply as heat. For our homogeneous and isotropic reference material, $s_{\rm c}$ has the same value for all bonds. The critical bond stretch can be related to the fracture energy $G$ (energy per unit fracture surface) by evaluating the energy needed to break all bonds across a unit area. The result is  \citep{bobaru2016handbook} 
\begin{equation}
    s_c = \left\{ \begin{array}{ll}
    \sqrt\frac{\pi G}{3k\delta}, \quad d=2,\\
    \sqrt\frac{5 G}{9k\delta}, \quad d=3.
    \end {array}\right.
    \label{eq:criticalStretch}
\end{equation}
This scaling has an important consequence when considering the convergence of the peridynamic model to the classical continuum in the limit $\delta \to 0$. There are two distinct ways to take this limit: We may either keep $s_{\rm c}$ fixed with the consequence that $G$ must go to zero in proportion with $\delta$. In this case, the system without cracks has a finite and fixed failure stress but becomes infinitely flaw sensitive. Conversely, where we keep $G$ fixed, the behavior in presence of cracks has a well defined $\delta\to 0$ limit but in this limit, the system without cracks is infinitely strong. When studying the effects of the peridynamic length scale $\delta$ on strength in presence of disorder, both limits will be considered.  

\subsection{Stochastic heterogeneity}

In an intermediately homogenized stochastic model, the local density $\rho(\Bx)$, bond micro-modulus $c(\Bx,\Bx')$, and critical bond stretch $s_{\rm c}(\Bx,\Bx')$ are all stochastic quantities which may depend in a random manner on the respective spatial coordinates. Given that bonds are extended spatial objects, the question arises how to relate their properties to the underlying spatial manifold. Several approaches are conceivable - for instance, one might introduce a field of local elastic moduli or local elastic compliances and obtain the bond modulus or compliance by averaging (i.e. integrating) this field along the bond. Here we adopt the method of introducing a field of local elastic moduli $\kappa(\Bx)$ and evaluating the bond modulus from the average of the elastic moduli at the bond endpoints,
\begin{equation}
c(\Bx,\Bx') = \frac{c(\Bx) + c(\Bx')}{2} \quad,
\label{eq:bondrand}
\end{equation}
where $c(\Bx)$ is calculated according to \autoref{eq:bondModulus}.
As fundamental fields in our statistical model, we thus use the stochastic fields $\psi(\Bx), \psi(\Bx) \in \{\rho(\Bx)$ (density), $\kappa(\Bx)$ (modulus), $s_{\rm c} (\Bx)$ ( failure strain)$\}$. These fields are, in general, mutually correlated and may possess a complex spatial correlation structure. Here we use a simplified statistical description where the fields $\kappa, s$ can be envisaged as monotonic functions of the local density $\rho$: $\kappa(\Bx) = \kappa(\rho(\Bx)), s(\Bx) = s(\rho(\Bx))$. The statistical properties of the field $\rho$ are characterized by the probability distribution function $F(\rho)$ and spatial pair correlation function $\Phi_{\rho}(\Bx,\Bx')$, while multi-point correlations are neglected. Moreover, we assume spatial homogeneity and isotropy, hence $\Phi(\Bx´,\Bx') = \Phi_{\rho}(\xi)$. The functions $\Phi_{\rho}(\xi)$ and $p(\rho)$ are constitutive for the statistical model. They need to be obtained from statistical analysis of experimental data, or by coarse-graining of lower-scale models. Moreover, by $\mu_{\rho}$ we denote the mean and by $\sigma_{\rho}$ the variance of the random field $\rho$. 

To numerically implement the model, we use a method discussed by \citet{vio2001numerical} following \citet{grigoriu2000non}. We construct an auxiliary field $\phi(\Bx)$ where $\phi$ is a  Gaussian random field with spatial correlation function $\Phi_{\phi}(\xi)$ and distribution function 
\begin{equation}
G(\phi) = \frac{1}{2}\left[1 + {\rm erf}\left(\frac{\phi}{\sqrt{2}}\right)\right]
\end{equation}.
For a Gaussian random field, the joint probability density function of the two variables $\phi_1 = \phi(\Bx)$ and $\phi_2 = \phi(\Bx + \Bxi)$ is given by
\begin{equation}
    p(\phi_1,\phi_2;\Phi_{\phi}(\xi)) = \frac{1}{2\pi \sqrt{1 - (\Phi_{\phi}(\xi))^2}} \exp\left[-\frac{\phi_1^2 + \phi_2^2 + 2 \phi_1 \phi_2 \Phi_{\phi}(\xi)}{2[1 - (\Phi_{\phi}(\xi))^2)]}\right]
\end{equation}

We now construct a mapping $\rho(\phi)$ that transforms the probability distribution function of $\phi$ to that of $\rho$
\begin{equation}
    \rho(\phi) = F^{-1}(G(\phi)).
    \label{eq:phitorho}
\end{equation}
The relationship between the spatial correlation functions of $\rho$ and $\phi$ is then given by 
\begin{equation}
    \Phi_{\rho}(\xi) = \frac{1}{\sigma_{\rho}^2 }
    \int \int [\rho(\phi_1) - \mu_{\rho}][\rho(\phi_2) - \mu_{\rho}] p(\phi_1,\phi_2;\Phi_{\phi}(\xi)) \diff \phi_1 \diff \phi_2
    \label{eq:corrtransgen}
\end{equation}
By inverting this relationship analytically or numerically, it is possible to construct, for a given correlation function of $\rho$, a matching correlation function for the auxiliary field $\phi$. One can then use standard methods such as the randomization method described by \cite{Hesse2014May} to construct the auxiliary field $\phi(\Bx)$, and finally use Eq. (\ref{eq:phitorho}) to transform to the sought field $\rho(\Bx)$. 

\subsubsection{Implementation for a model material}

As a model material, we consider a random porous material with log-normal distribution of the local density and exponentially decaying correlations of the auxiliary field $\phi$. The mapping between the auxiliary field $\phi$ and the density field $\rho$ is assumed as
\begin{equation} 
\rho(\Bx) = \rho_0 \exp\left[ \alpha_{\rho} \left(\phi(\Bx) - \frac{a}{2}\right)\right] 
\label{eq:rhophi}
\end{equation}
where $\rho_0$ is the mean density and the parameter $\alpha_{\rho}$ controls the scatter of local densities. Typical for random porous materials, the local modulus and strength are considered as powers of the local density \citep{gibson1982mechanics}. Thus, $\kappa = \kappa_0 (\rho/\rho_0)^g$, $s_{\rm c} = s_0 (\rho/\rho_0)^h$ where $g,h$ are Gibson-Ashby exponents for modulus and strength and $\kappa_0,s_0$ are the characteristic modulus and failure strain of the porous material.

Combining these relations with Eq. (\ref{eq:rhophi}) we obtain a direct mapping between the auxiliary field $\phi$ and the parameter fields $\kappa,s_{\rm c}$:
\begin{equation} 
\kappa(\Bx) = \kappa_0 \exp\left[ \alpha_{\kappa} \left(\phi(\Bx) - \frac{\alpha_{\kappa}}{2}\right)\right] \quad,\quad 
s(\Bx) = s_0 \exp\left[ \alpha_{s} \left(\phi(\Bx) - \frac{\alpha_s}{2}\right)\right].
\label{eq:kappas}
\end{equation}
The correlation function of the auxiliary field is taken as 
\begin{equation}
\Phi_{\phi}(\xi) = \exp\left[- \frac{\xi}{\lambda}\right]
\label{eq:phicorrexp}
\end{equation}
where $\lambda$ is a correlation length.

In the following we consider fluctuations in local modulus only, i.e., we set $\alpha_{s} = \alpha_{\rho}= 0, \alpha_{\kappa} = \alpha$. The probability density of the local modulus $\kappa$ is with \autoref{eq:kappas} given by
\begin{equation}
p(\kappa) = \frac{1}{\alpha \kappa \sqrt{2\pi}}
\exp\left[-\frac{1}{2 \alpha^2} \left(\ln\frac{\kappa}{\kappa_0} -\frac{\alpha^2}{2}\right)^2\right]
\end{equation}
with the mean, variance, and coefficient of variation given by 
\begin{equation}
    \langle \kappa \rangle = \kappa_0\;,\quad
    \sigma^2_{\kappa} = \kappa_0^2 [\exp \alpha ^2 - 1]\;,\quad
    CV_{\kappa} = \sqrt{\exp \alpha^2 - 1} \;.
\end{equation}
The correlation function of the local modulus fluctuations as obtained from \autoref{eq:phicorrexp} and \autoref{eq:corrtransgen} is then given by the Gumbel-like expression \citep{vio2001numerical}
\begin{equation}
\Phi_{\kappa}(\xi) = \frac{\exp\left[\alpha^2\exp\left(- \frac{\xi}{\lambda}\right) \right] -1}{\exp\left[ \alpha^2\right] - 1}.
\end{equation}

Moreover, we set the default value of the correlation length equal to the peridynamic horizon, $\lambda = \delta$, reflecting the idea that the peridynamic horizon and the fluctuation correlation length may represent the same microstructural length scale. Situations where $\delta$ and $\xi$ are varied independently are considered later in Section 3.3.2.  

Regarding material parameters, we use mechanical properties of snow as a example of a highly porous and disordered material as reported in \citet{VanHerwijnen2016Dec}, in which disorder has a significant impact on fracture properties and may even influence avalanche risk \cite{fyffe2004effects}. The high degree of disorder that may be associated with the porous microstructure of snow is manifest by the scatter of strength data as reported by \citet{kirchner2004size} whose data for the failure stress of homogeneous snow samples in bending tests show a coefficient of variation $CV \approx 0.75 \pm 0.25$. Accordingly, we shall focus on moderate to large degrees of disorder with $\alpha$ values leading to comparable variation coefficients. Moreover, we set the default value of the horizon to represent the typical scale of a coarse-grained snow microstructure, $\delta=4$ mm. 
\begin{table}[h]
    \centering
    \begin{tabular}{c c c}
      $\rho$   &  $E$   &   $G_{\rm c}$ \\
        kg/m$^3$ & MPa & J/m$^2$\\ \hline
        200 & 3 & 0.4
    \end{tabular}
    \caption{Material parameters used in simulations}
    \label{tab:materialParameters}
\end{table}

With the parameters compiled in \autoref{tab:materialParameters}, Equation (\ref{eq:criticalStretch}) then gives us a critical bond stretch $s_c=0.007$. Finally, we note that in our numerical implementation we consider the case $d=2$, which in bond-based peridynamics implies a Poisson ratio of $\nu = \frac{1}{3}$.

\subsection{Numerical aspects}

As default, we discretize the continuous medium using a regular grid of collocation points with spacing $\Delta$. Using a  regular grid may lead to convergence issues owing to the problem of inaccurate estimation of a lattice cell volume intersected by a spherical neighbourhood. We employ the algorithm suggested by \cite{Scabbia2023Feb}  to address this issue. The numerical implementation converges to the limit of \autoref{eq:forcebalance} as $m = \delta/\Delta \to \infty$ (so-called $m$ convergence). 

We use the python library GSTools \citep{Muller2022Apr} to generate an ensemble of 2D Gaussian random fields $\phi(\Bx)$ with the correlation function (\ref{eq:phicorrexp}). The random fields $\phi(\Bx)$ are then used to create a field of bulk moduli, which are in turn used to evaluate bond moduli following Eq. (\ref{eq:bondrand}). As default, we simulate 10 realizations of the random field for each set of parameters. We implement the interemediately homogenized stochastic model in Peridigm \citep{littlewood-2023}. 

\section{Simulation results}
\subsection{Simulations of the model material}

We study general aspects of the proposed modelling approach by simulations of the random model material described in Section 2.2. This study comprises an investigation of the effects of discretization and the convergence of the model to the continuum limit. 

\subsubsection{m-convergence}
In order to test the mesh sensitivity of the model, we perform an m-convergence study, where $m=\delta/\Delta$ is increased while keeping the horizon $\delta$ fixed. For a classical bond-based peridynamic setup without disorder, an m-convergent model is expected to converge to the non-local solution for given $\delta$. For the present IH model, we also expect the disorder effects to be independent of the value of $m$ or the discretization, at least for sufficiently large $m$.

We carry out a series of simulations of samples of size $ L \times L$ where $L = 200$mm. A central notch of length $80$mm is created along the plane $y = L/2$ by using a bond filter that deletes all the bonds crossing the plane. The horizon is fixed at $\delta = 4$mm. Displacements are imposed on two boundary layers of thickness $\delta$ at the top and and bottom surfaces $y=0$ and $y=L$. The boundary layer at $y=0$ is kept fixed while the layer at $y=L$ is displaced in y-direction according to the following scheme:
\begin{equation}
    u_y(y=L) = \begin{cases}
        \frac{v}{2t_a}t^2, & \text{if $t\leq t_a$} \\
        v(t-\frac{t_0}{2}), & \text{if $t>t_a$,}
    \end{cases}
\end{equation}
where $u_y(y=L)$ is the displacement of the boundary layer, $v$ is the displacement rate, and $t_a$ is an acceleration period. 

\begin{figure}
    \centering
    \includegraphics[width=\textwidth]{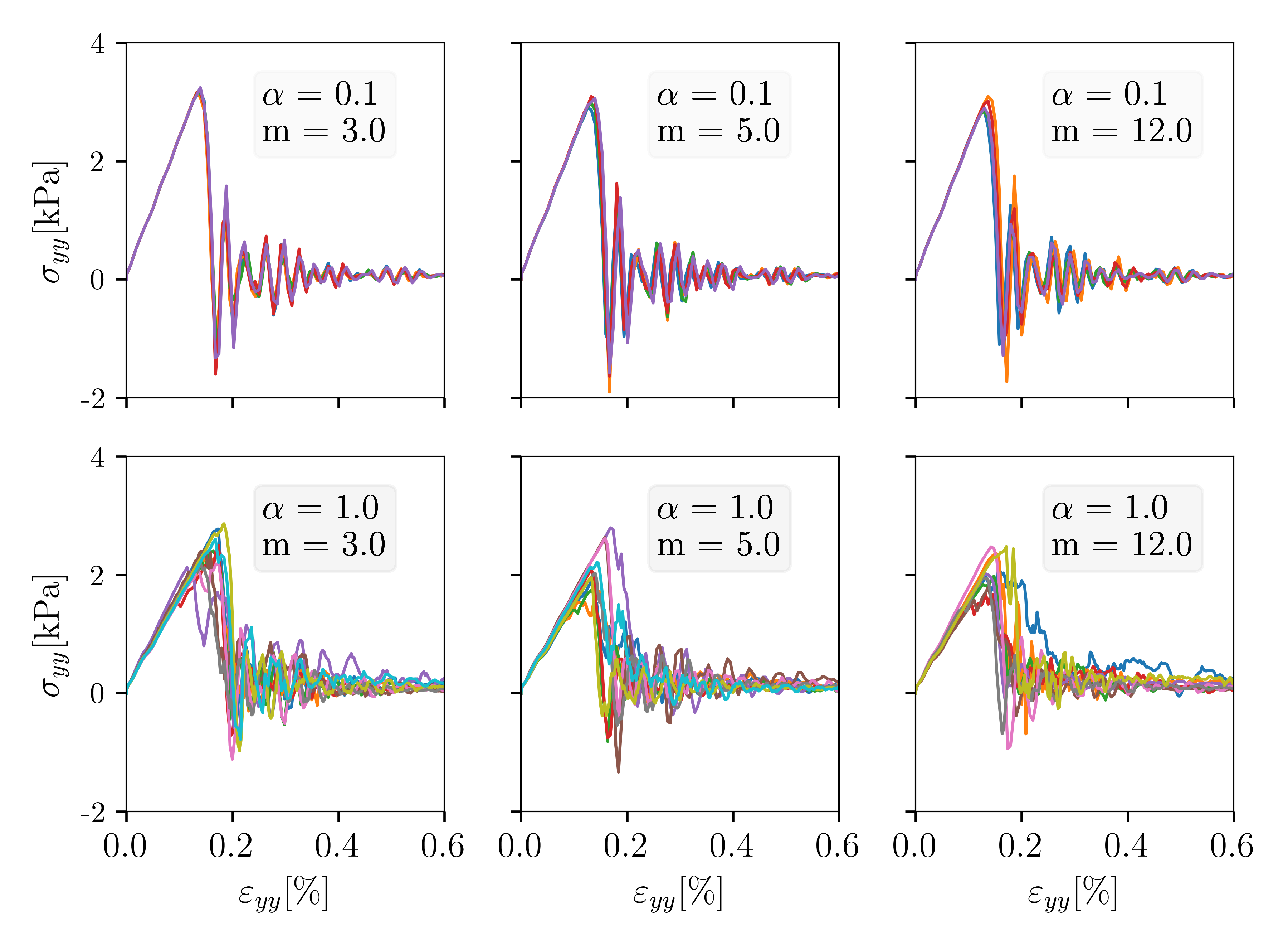}
    \caption{Stress-strain curves of uniaxial tensile simulations for different values of $m$ and $\alpha$. Rows: $\alpha=0.1$(top) and $\alpha=1.0$(bottom); columns: $m= 3.0,5.0,12.0$(left to right). Different colours indicate samples with different random fields.}
    \label{fig:stressstraincurves}
\end{figure}

\begin{figure}[htb]
    \includegraphics[width=\textwidth]{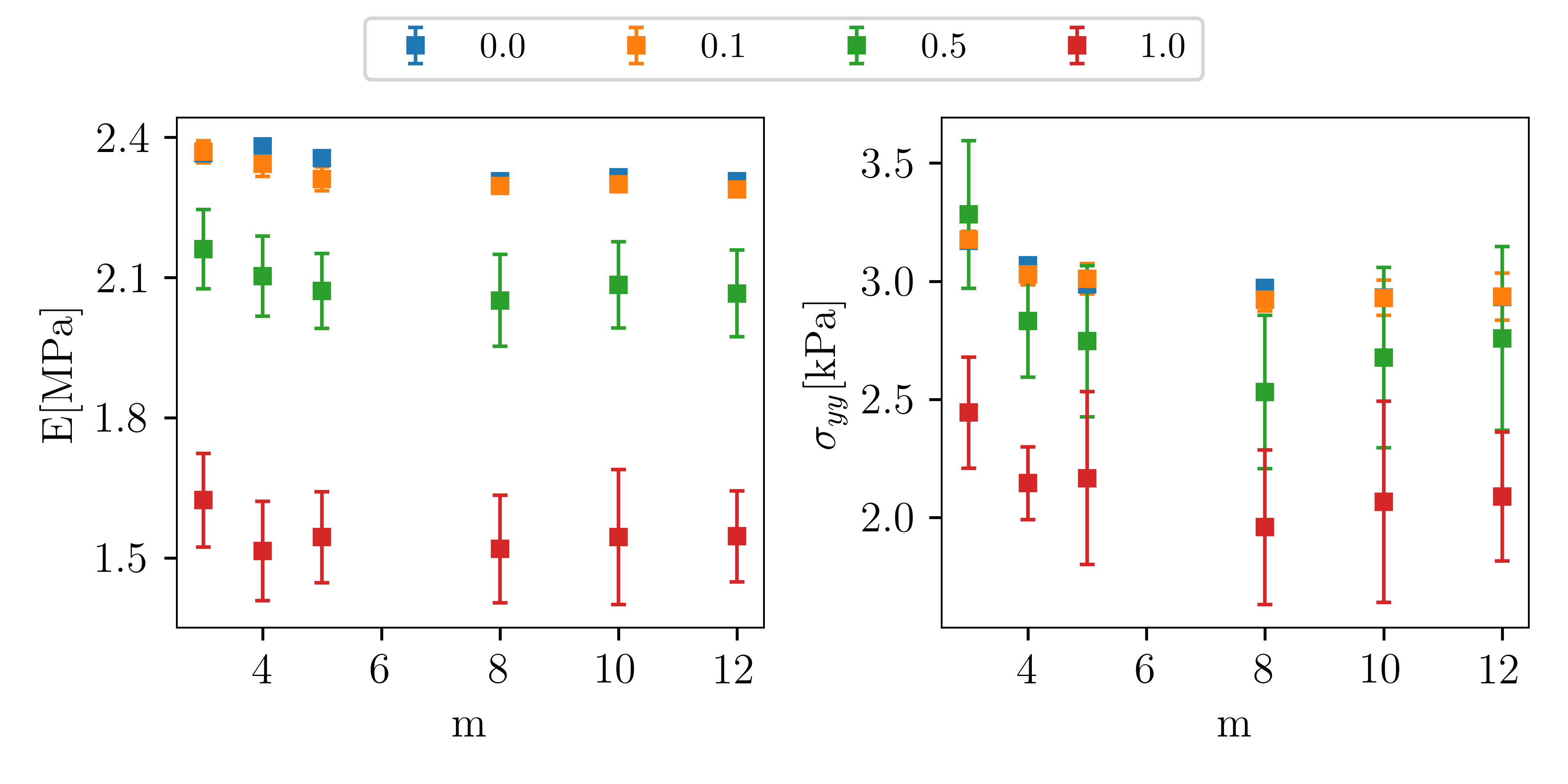}
    \caption{Effective elastic modulus and peak stress as functions of $m$ for different degrees of disorder. The square points represent mean values evaluated from 10 samples and the error bars show the corresponding standard deviation. }
    \label{fig:mConvergence}
\end{figure}

We control the disorder within the system by increasing the value of the scaling parameter $\alpha$ and perform, for each value of $\alpha$, a series of simulations with different grid spacing $\Delta$. Results of the simulated tensile tests are shown in \autoref{fig:stressstraincurves}. From the stress-strain curves, we extract the initial linear slope before the onset of damage, which define an effective elastic modulus of the cracked sample, and the peak stress. These parameters are plotted in \autoref{fig:mConvergence} as functions of $m$.

For increasing disorder parameter $\alpha$, the samples become weaker as evident by a decrease in both elastic modulus and peak stresses. At the same time, the elastic modulus and the peak stress are, within the statistical scatter evaluated as the variance over 10 samples, almost independent of the value of $m$ for $m>3$. This behavior does not depend appreciably on the degree of disorder: For all investigated $\alpha$ values in the range $0 \le \alpha \le 1$, corresponding to coefficients of variation of the local elastic modulus distribution in the range $0 \le {\rm CoV} \le 1.35$, we find that mean values are independent on $m>3$ within the statistical variation of the results, and that this variation does not systematically depend on $m$. Thus, for $m>3$ the statistics of sample behavior shows no appreciable dependence on discretization. 

\subsection{Damage patterns}

\begin{figure}[b]
    \includegraphics[width=\textwidth]{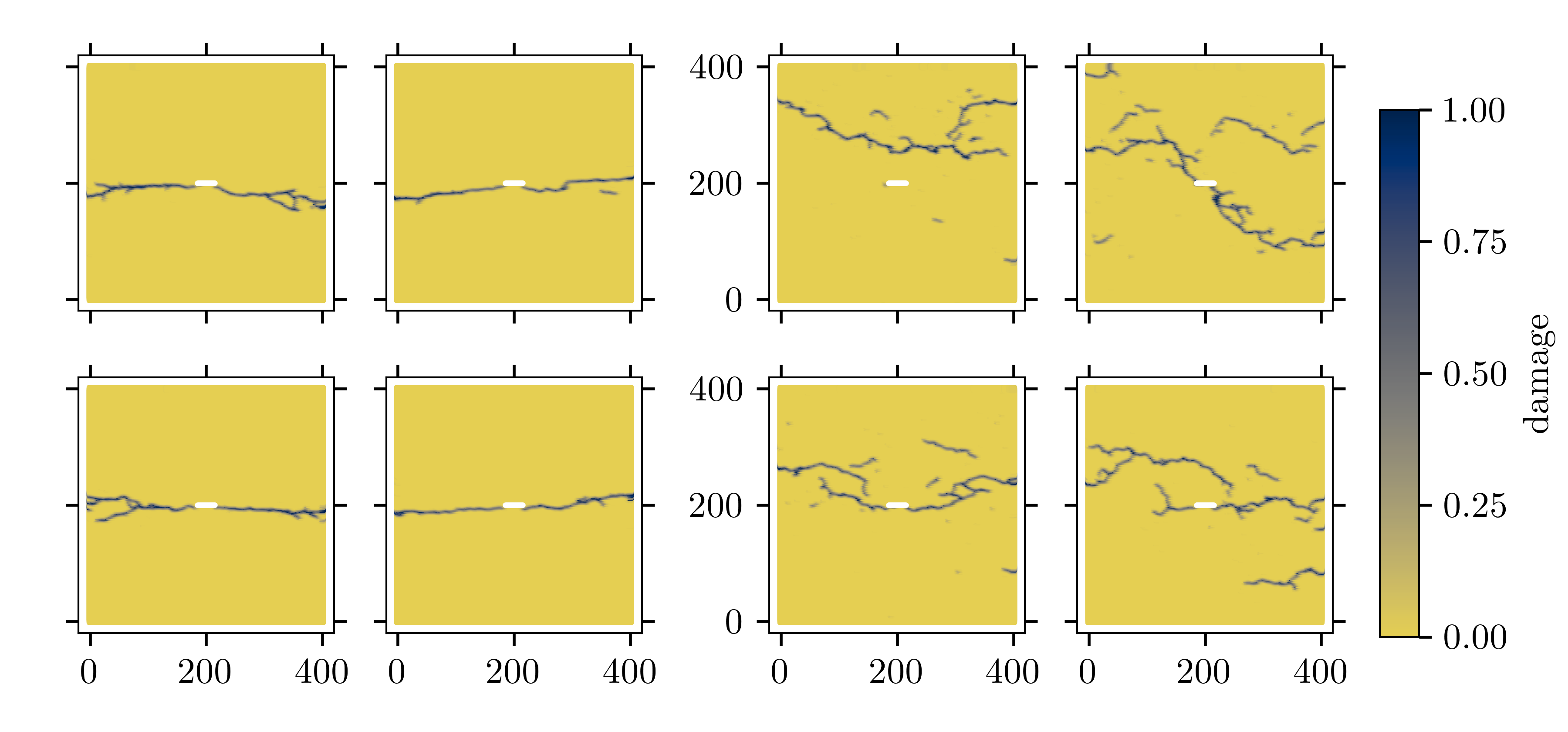}
    \caption{Comparison of damage patterns, system size $400 \times 400$ mm, $\delta = \xi = 4$mm, initial crack length $a=16$ mm, left 4 samples: $\alpha = 0.5$, right 4 samples: $\alpha = 1$.}
    \label{fig:damagepattern}
\end{figure}

The patterns of damage accumulation are strongly influenced by the degree of disorder. We illustrate this by plotting spatial patterns of the damage parameter $D(\Bx)$, defined as the fraction of broken bonds among all bonds connecting to the material point $\Bx$ in the reference configuration. We plot $D(\Bx)$ at $2.4$mm displacement of the upper boundary layer, at which macroscopic cracks have already propagated through all of the samples. Results are shown in \autoref{fig:damagepattern} for samples of size $L=400$mm containing a load-perpendicular initial crack of length $2a=32$mm.

For low disorder, we generally observe crack patterns that follow the deterministic prediction, i.e., for a load-perpendicular initial crack under uniaxial tensile loading, the crack propagates in pure mode I. 

At intermediate disorder ($\alpha = 0.5$, \autoref{fig:damagepattern}, left) we find small deviations from model-I cracking as well as occasional, disorder-induced crack branching while damage still accumulates by extension and branching of the pre-existing crack. 

At high disorder ($\alpha = 1$, \autoref{fig:damagepattern}, right) we observe extensive crack branching concomitant with profuse crack nucleation at locations away from the initial, pre-existing crack. In some realizations we find that the initial crack does not propagate at all and failure is controlled by the propagation of a crack that nucleates spontaneously at a different location in the simulated sample. The very wide spectrum of damage patterns observed in different realizations illustrates the necessity for a non deterministic description of the failure process in order to screen the range of possible failure scenarios. 

\subsection{Size Effect}

We now investigate the stochastic size effect captured by the intermediately homogenized model. To this end, we study 2D square samples of sides 50 mm, 100 mm, 200 mm and 400 mm. We first study samples without pre-cracks and choose $m=4.0$ and $\delta=4$ mm. The loading protocol remains the same as the previous sections, as does the ensemble size for the random fields. 

\begin{figure}[bth]
    \includegraphics[width=\textwidth]{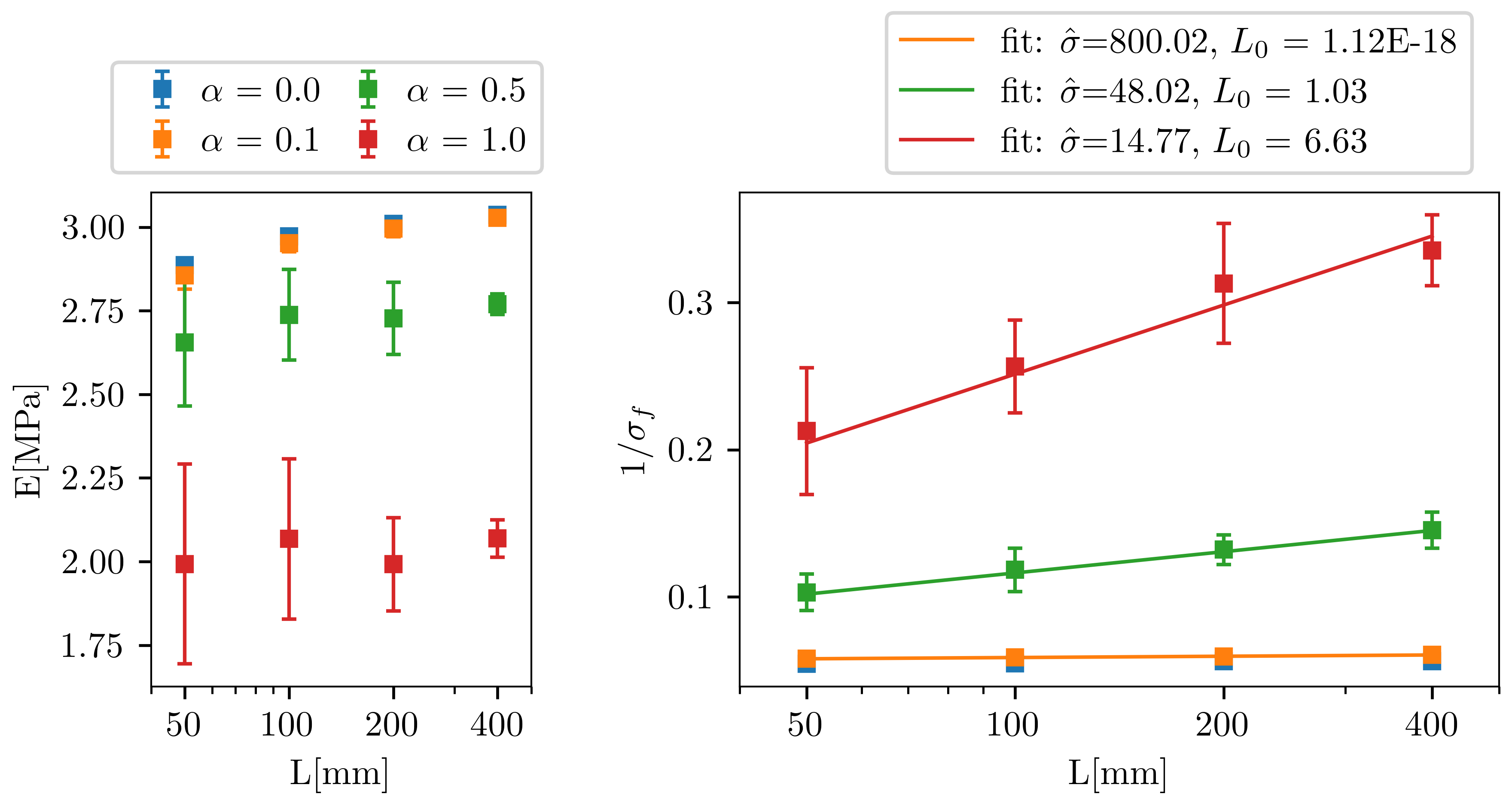}
    \caption{Elastic response and peak stress of unnotched samples with different sizes and different degrees of disorder; left: apparent elastic modulus (initial slope of stress-strain curve), right: peak stress. }
    \label{fig:sizeEffect}
\end{figure}

\autoref{fig:sizeEffect}, left, shows that for smaller sample sizes, the elastic response of the material is affected by the peridynamic surface effect, and the apparent elastic modulus is therefore lower. For high disorder, $\alpha = 1$, the large scatter of the apparent moduli renders the effect statistically insignificant. 

Regarding peak strength of samples without macroscopic cracks, we observe that the strength of disordered samples decreases with increasing system size. This decrease can be described by a relationship derived first by \citet{duxbury1987breakdown} in the context of breakdown of random fuse networks, and later found in a wide range of disordered materials in the regime of sufficiently large system sizes $L$:
\begin{equation}
    \sigma_{\rm f} = \frac{\hat{\sigma}}{1 + \log (L/L_0)}.
    \label{eq:dux}
\end{equation}
This implies that a plot of $(1/\sigma_{\rm f})$ versus $L$ produces a straight line. Our data are consistent with this idea as illustrated in \autoref{fig:sizeEffect}, right. We note that data with low or without disorder cannot be meaningfully described by Eq \ref{eq:dux} as seen from the physically meaningless value of the fit parameters $\hat{\sigma},L_0$ for samples with $\alpha = 0$ and $\alpha = 0.1$. These data are best described as $L$ independent. 

In presence of pre-existing macroscopic cracks, the size dependence of the failure stress is controlled by the interplay between statistical effects and fracture mechanics considerations. Since our intermediately homogenized model contains two internal length scales, the peridynamic horizon $\delta$ and the correlation length $\lambda$, as well as the disorder parameter $\alpha$, it is worthwhile exploring how the interplay between fracture mechanics and statistical effects depends on these parameters. To this end, we have conducted extensive simulations varying the different parameters. In order to analyze the dependency of failure stress on crack length in our simulations, we use a phenomenological relation originally proposed by \citet{mcclintock1965plasticity} in the context of elastic-plastic fracture: 
\begin{equation}
    \sigma_{\rm f} = \frac{K_{\rm c}}{\sqrt{\pi(a_0+a)}}.
    \label{eq:Irwin}
\end{equation}
Here $K_c$ is the effective fracture toughness of the material, $a$ is the length of the initial crack, and $a_0$ is a characteristic length which may be understood as the size of the fracture process zone. 

\subsubsection{Dependence of strength on system size and disorder}
\begin{figure}[htb]
\centering
    \includegraphics[width=\textwidth]{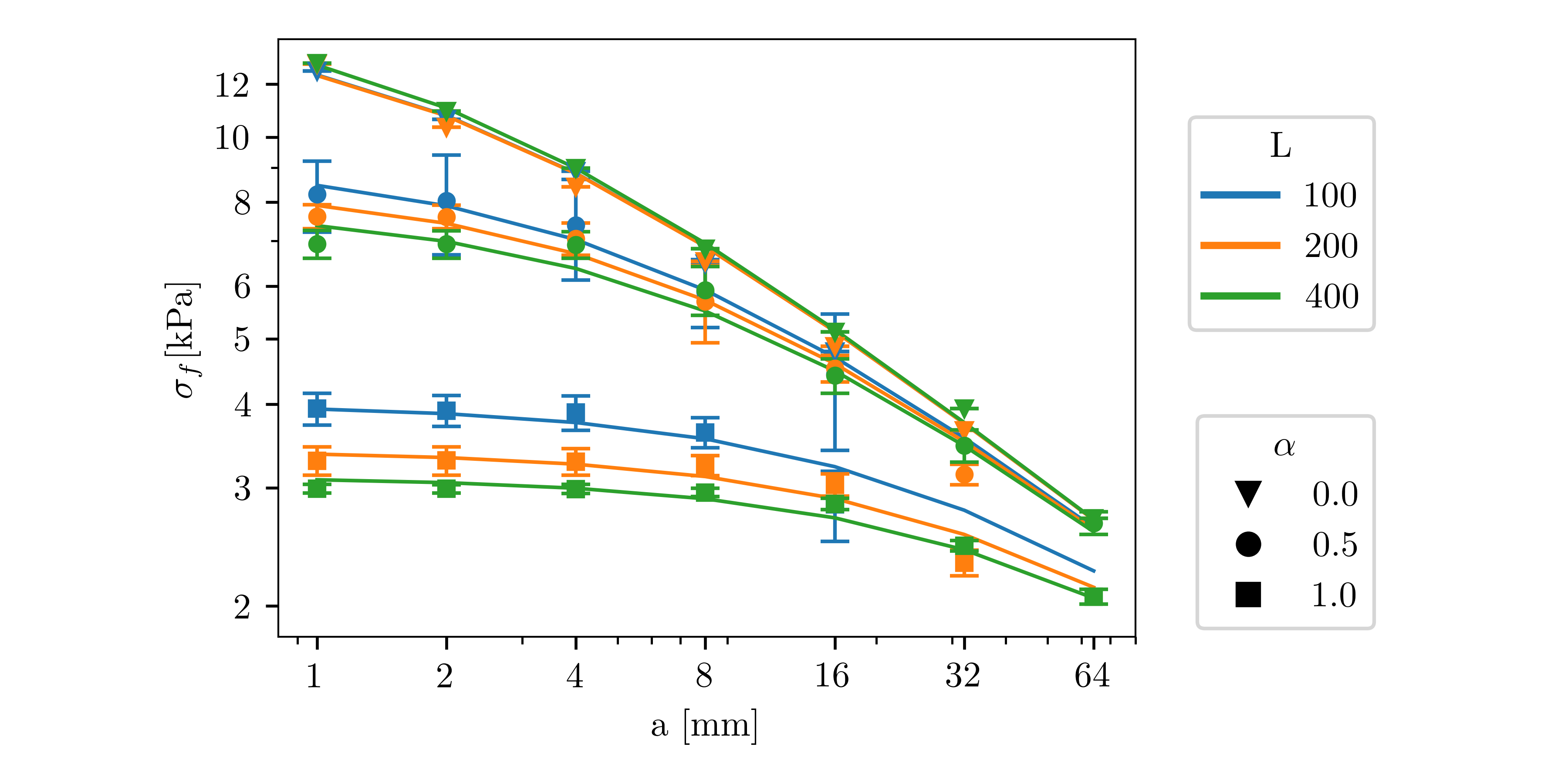}
    \caption{Failure stresses for varying system size $L$ and initial notch length $a$, for other parameters see text; the fit curves were obtained from \autoref{eq:Irwin} using the parameters of statistical model H3 in \autoref{tab:fitsAlphaL}.}
    \label{fig:fitsL}
\end{figure}

First we focus on the dependence of the parameters $K_{\rm c}$ and $a_0$ on system size $L$ and disorder parameter $a$. We study 2D square samples of edge length $L  \in $ [100mm, 200mm, 400mm], containing central load-perpendicular notches with half-length $a \in $[1mm, 2mm, 4mm, 8mm, 16mm, 32mm, 64mm]. Only specimen with $L/a > 5$ were considered to ensure that the stress intensity factors are not noticeably affected by boundary conditions at the specimen edges.  The simulated microstructures had disorder parameters  $\alpha \in [0, 0.5, 1]$. In all simulations, the length scale parameters were set to $\delta = \lambda = 4$mm. For each set of parameters, 10 realizations of the disordered microstructure were simulated. The resulting failure stresses and their standard deviations are compiled in \autoref{fig:fitsL}.

\begin{figure}[htb]
    \begin{minipage}{.49\textwidth}
    \includegraphics[width=\textwidth]{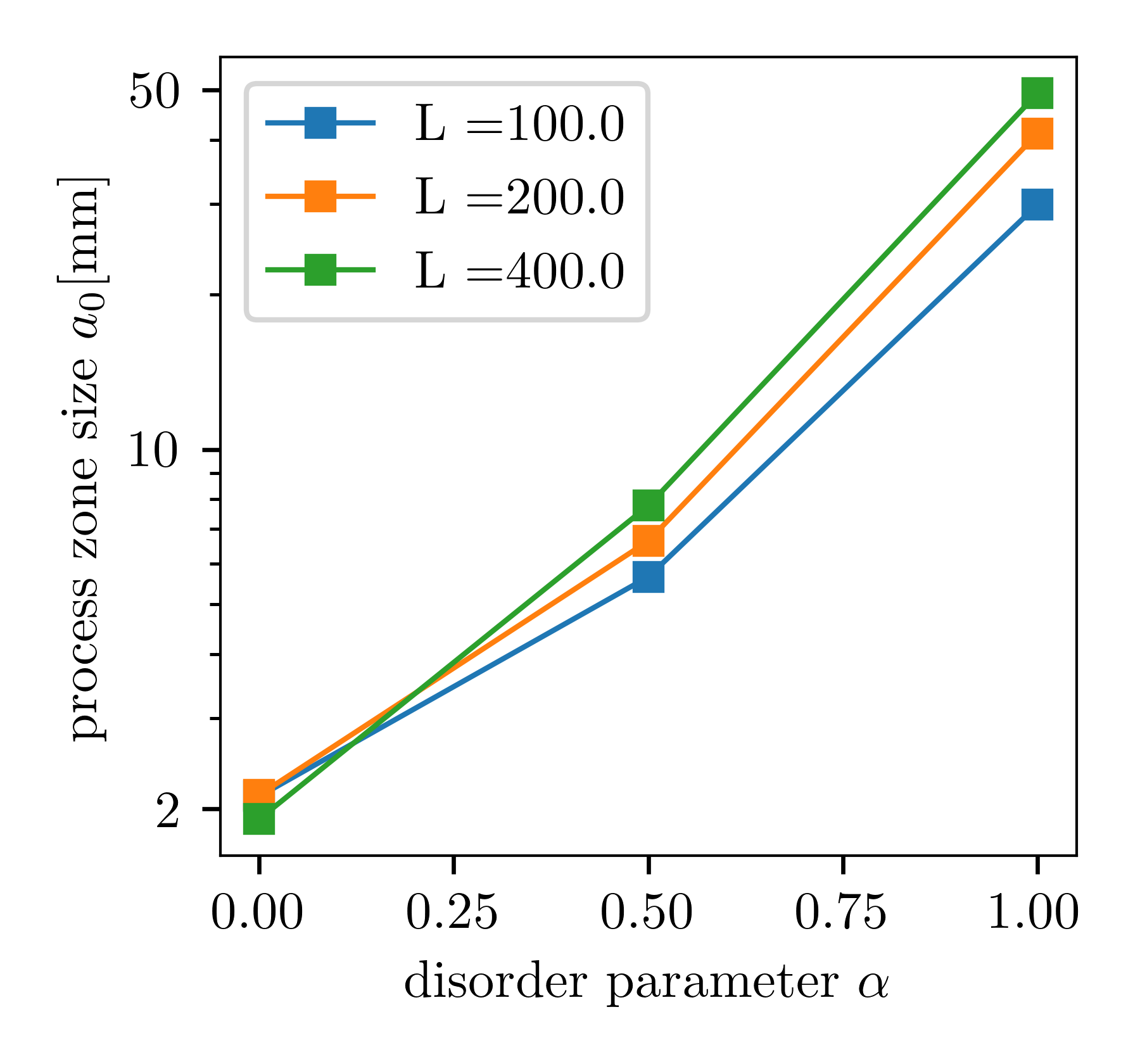}
    \end{minipage}
    \hfill
    \begin{minipage}{.49\textwidth}
    \includegraphics[width=\textwidth]{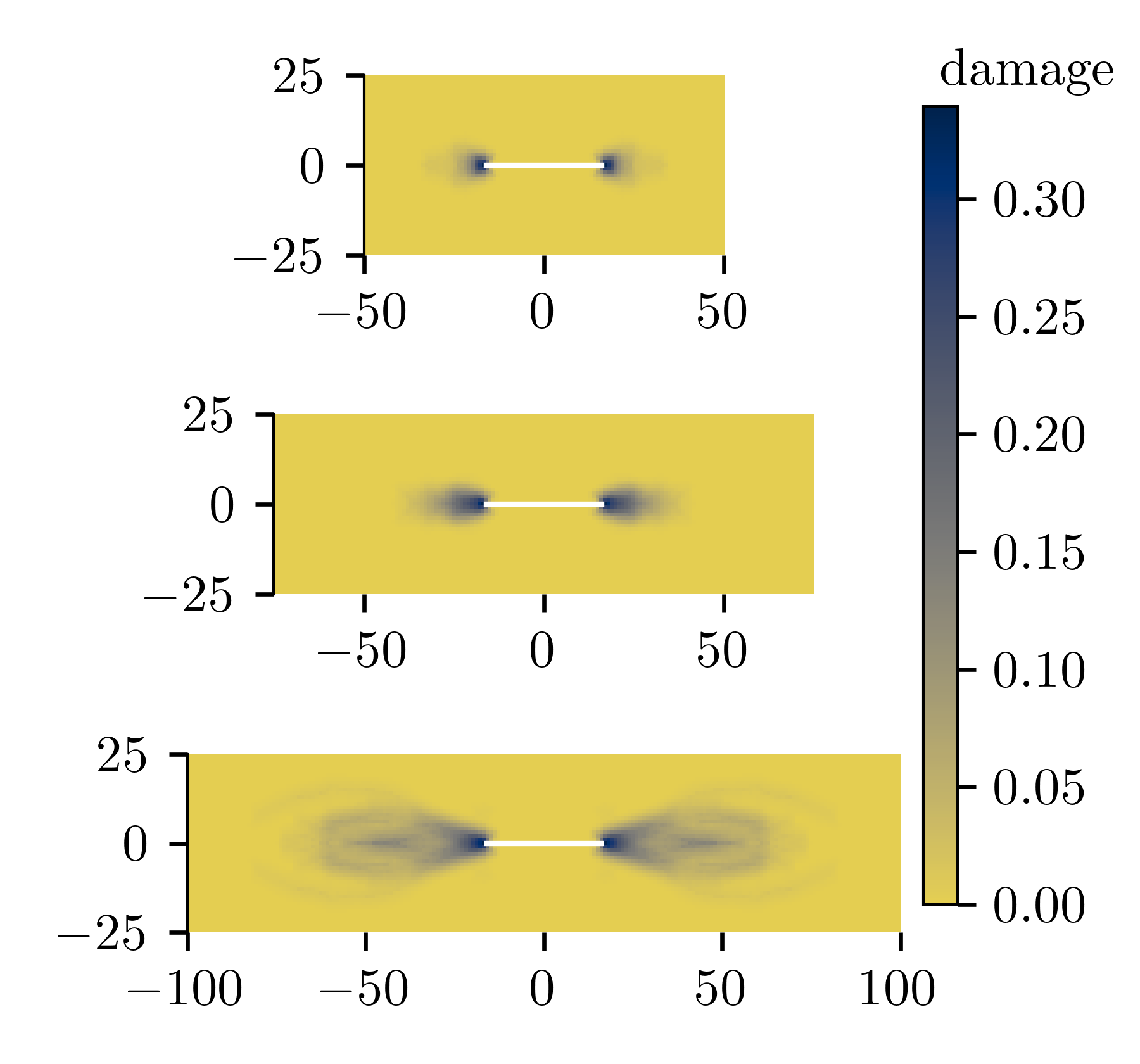}
    \end{minipage}
    \caption{Left: dependency of process zone size $a_0$ on disorder parameter $\alpha$ for different system sizes; right: damage distributions around the tip of the initial crack at the peak stress, for $\alpha = 0.5$ and different values of $L$, the local damage variables have been averaged over 10 realizations of the microstructure and over 4 symmetry equivalent configurations for each realization.}
    \label{fig:a0vs_alpha_L}
\end{figure}

To analyze the data, we formulate 3 statistical models as follows. H1: Both $K_c$ and $a_0$ depend on both $\alpha$ and on $L$; H2: $K_c$ depends on $\alpha$ and $a_0$ depends on both $\alpha$ and on $L$; H3: $K_c$ is the same for all samples but $a_0$ depends on both $\alpha$ and $L$. Other  hypotheses (e.g. independence of $a_0$ on $\alpha$ and/or on $L$) could be ruled out by direct inspection of the data. 
We then tested hypotheses H1-H3 using the Aikake information criterion as described by \citet{Anderson1998-ur},
\begin{equation}
    AIC = 2K - 2\ln\cal{L},
    \label{eq:akaike}
\end{equation}
where K is the number of fit parameters and $\cal{L}$ is the obtained maximum value of the likelihood function, evaluated under the assumption of Gaussian statistics. Results are compiled in \autoref{tab:fitsAlphaL}. From the aforementioned models, H3 gives us the lowest value of AIC, which implies that our results are statistically best described by a fracture toughness that is independent on system size and disorder, but a process zone size $a_0$ that depends significantly on both parameters. The resulting dependencies are shown in \autoref{fig:a0vs_alpha_L}, left. 
\begin{table}[h]
    \centering
    \begin{tabular}{|*{9}{c|}}
    \hline
    \multirow{3}{3em}{Model} & \multirow{3}{3em}{AIC} & \multirow{3}{3em}{$\alpha$} & \multicolumn{6}{|c|}{L} \\ \cline{4-9}
      &  & & \multicolumn{2}{|c}{100 mm} & \multicolumn{2}{|c|}{ 200 mm} & \multicolumn{2}{c|}{400 mm} \\ \cline{4-9}
      & & & $K_c$ & $a_0$ & $K_c$ & $a_0$ & $K_c$ & $a_0$ \\ \hline 
      \multirow{3}{3em}{H1} & \multirow{3}{3em}{-361.7} & 0 & 36.4 & 1.7 & 35.3 & 1.5 & 38.5 & 1.9 \\ \cline{3-9}
      & & 0.5 & 38.6 & 5.6  & 37.9 & 6.2 &  41.5 & 9.4\\ \cline{3-9}
      & & 1.0 & 30.8 & 17.5 & 36.5 & 36.0 & 41.5 & 57.8  \\ \hline
      \multirow{3}{3em}{H2} & \multirow{3}{3em}{-367.1} & 0 & 36.9 & 1.8 & 36.9 & 1.8 & 36.9 & 1.6 \\ \cline{3-9}
      & & 0.5 & 39.4 & 5.9  & 39.4 & 7.0 &  39.4 & 8.1 \\ \cline{3-9}
      & & 1.0 & 36.7 & 26.4 & 36.7 & 36.3 & 36.7 & 43.5  \\ \hline
      \multirow{3}{3em}{H3} & \multirow{3}{3em}{-368.8} & 0 & 38.7 & 2.1 & 38.7 & 2.1 & 38.7 & 1.9 \\ \cline{3-9}
      & & 0.5 & 38.7 & 5.7  & 38.7 & 6.6 &  38.7 & 7.8 \\ \cline{3-9}
      & & 1.0 & 38.7 & 29.9 & 38.7 & 41.1 & 38.7 & 49.2  \\ \hline
    \end{tabular}
    \caption{Fit results of the study with different L and $\alpha$}
    \label{tab:fitsAlphaL}
\end{table}

For zero disorder, the process zone size $a_0 \approx 2$mm is of the order of the internal length scales $\delta = \lambda$. However, for disordered materials the values of $a_0$ are significantly larger and depend on the size of the simulated samples in a statistically significant manner. The idea of a process zone size that depends on the size of the simulated sample is borne out by the simulated damage distributions, see \autoref{fig:a0vs_alpha_L}, right. 

\subsubsection{Dependence of strength on internal length scales}

The following studies focus on the question how $K_{\rm c}$ and/or $a_0$ are influenced by the two length scales present in our model, the horizon $\delta$ and the correlation length of the random field $\lambda$. First, we consider situations where $\delta$ and $\lambda$ are equal. 

As explained in Section 2.1, changing the horizon according
to \autoref{eq:criticalStretch} implies changing either the critical strain $s_{\rm c}$ or the fracture energy $G$. From \autoref{eq:Irwin} together with \autoref{eq:criticalStretch} and $K_{\rm c} \approx \sqrt{GE}$ we find
\begin{equation}
    \sigma_{\rm f} = \sqrt{\frac{GE}{\pi(a_0+a)}}. 
    \label{eq:Irwin2}
\end{equation}
Keeping the critical stretch fixed when changing $\delta$ will change the fracture energy $G$ and thereby the asymptotic behavior at large $a$, whereas keeping the fracture energy fixed will according to \autoref{eq:criticalStretch} change $s_{\rm c}$ and thereby the asymptotic behavior in the crack-free limit. We consider both scenarios: (a) scaling the critical stretch to keep the fracture energy fixed, (b) keeping the critical stretch unchanged, which implies to increase the fracture energy in proportion with $\delta$. 

\begin{figure}[htb]
    \centering
    \includegraphics[width=\textwidth]{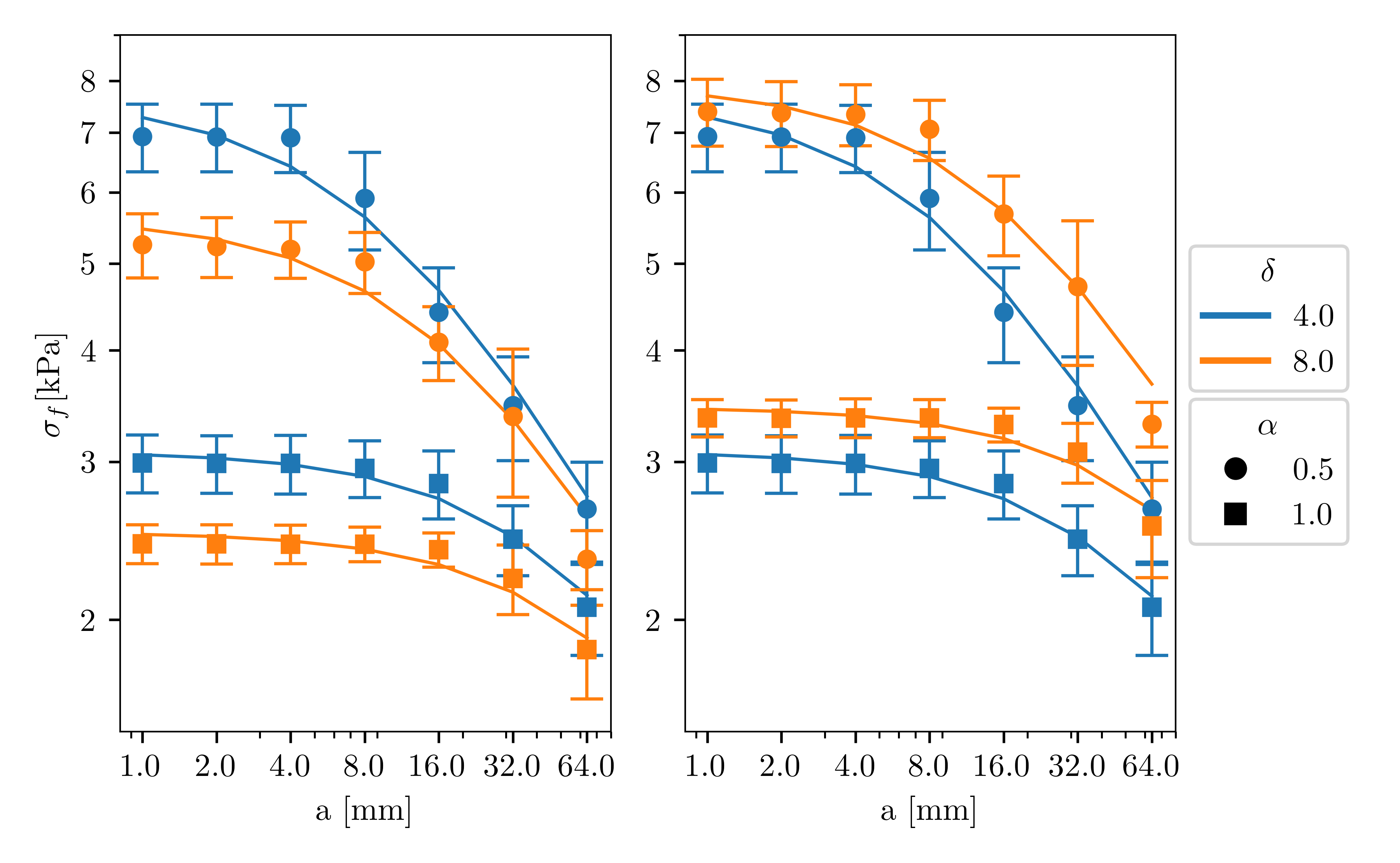}
    \caption{Failure stress vs notch half length: dependency on disorder parameter and horizon; all simulations were conducted with $L = 400$mm; left: simulations with different values of $\lambda = \delta$ and fixed nominal fracture energy, fit functions obtained from statistical model aH2; right: simulations with different values of $\lambda = \delta$ and fixed critical stretch, obtained from model bH2}
    \label{fig:fitsLengthScale}
\end{figure}

For scenario (a), we performed simulations with L = 400 mm, $\alpha = 0.5,1.0$, $\delta$ = 4 mm ($s_c$ = 0.007) and $\delta$ = 8 mm (re-scaled $s_c = 0.0049$). The corresponding simulation data are compiled in \autoref{fig:fitsLengthScale}, left.

For analyzing the failure stress data in \autoref{fig:fitsLengthScale}, left, the following statistical hypotheses are tested: (aH1) the data are best described by a $K_c$ that depends on $\delta$, and $a_0$ depends on ($\delta$,$\alpha$); (aH2) the data are represented by a value of $K_c$ that is independent of ($\delta$,$\alpha$), while $a_0$ varies significantly with ($\delta$,$\alpha$). Fit results are compiled in \autoref{tab:fitsGFixed}. 
\begin{table}[h]
    \centering
    \begin{tabular}{|*{7}{c|}}
    \hline
    \multirow{3}{3em}{Model} & \multirow{3}{3em}{AIC} & \multirow{3}{3em}{$\alpha$} & \multicolumn{4}{|c|}{$\delta$} \\ \cline{4-7}
      &  & & \multicolumn{2}{|c}{4 mm} & \multicolumn{2}{|c|}{8 mm} \\ \cline{4-7}
      & & & $K_c$ & $a_0$ & $K_c$ & $a_0$ \\ \hline 
      \multirow{2}{3em}{aH1} & \multirow{2}{3em}{-501.1} & 0.5 & 41.6 & 9.4 & 42.1 & 18.0 \\ \cline{3-7}
      & & 1.0 & 41.6 & 57.8 & 42.1 & 90.2 \\ \hline
       \multirow{2}{3em}{aH2} & \multirow{2}{3em}{-503.1} & 0.5 & 41.7 & 9.5 & 41.7 & 17.6 \\ \cline{3-7}
      & & 1.0 & 41.7 & 58.4 & 41.7 & 88.5 \\ \hline     
    \end{tabular}
    \caption{Fit results of the study with scaled $\delta$ and fixed fracture energy G}
    \label{tab:fitsGFixed}
\end{table}
Analysis using the Aikake information criterion indicates that aH2 best describes the data set, i.e., $K_c$ is independent of the length scales in the model, whereas $a_0$ increases significantly when increasing the internal length scale $\delta = \lambda$. 

Next we analyze scenario (b), performing simulations with L = 400 mm, $\alpha = 0.5,1.0$, $\delta = \lambda$ = 4 mm and 8 mm, while keeping $s_c$ fixed at 0.007 for both horizon sizes. The corresponding simulation data are compiled in \autoref{fig:fitsLengthScale}, right. For analyzing the data we test the following hypotheses:  
(bH1) The data are best described by a $K_c$ that are independently fitted for both values of $\delta$, and $a_0$ depends on ($\delta$,$\alpha$); (bH2) the data are best described by a single fit parameter $\tilde{K}$ from which the values of $K_c$ are evaluated as $K_{\rm c} = \tilde{K}\sqrt{\delta}$, while $a_0$ depends on both ($\delta$,$\alpha$). Fit results are compiled in \autoref{tab:fitsGScaled}. 

\begin{table}[h]
    \centering
    \begin{tabular}{|*{7}{c|}}
    \hline
    \multirow{3}{3em}{Model} & \multirow{3}{3em}{AIC} & \multirow{3}{3em}{$\alpha$} & \multicolumn{4}{|c|}{$\delta$} \\ \cline{4-7}
      &  & & \multicolumn{2}{|c}{4 mm} & \multicolumn{2}{|c|}{8 mm} \\ \cline{4-7}
      & & & $K_c$ & $a_0$ & $K_c$ & $a_0$ \\ \hline 
      \multirow{2}{3em}{bH1} & \multirow{2}{3em}{-417.0} & 0.5 & 41.6 & 9.4 & 58.7 & 17.5 \\ \cline{3-7}
      & & 1.0 & 41.6 & 57.8 & 58.7  & 91.8 \\ \hline
       \multirow{2}{3em}{bH2} & \multirow{2}{3em}{-419.0} & 0.5 & 20.76*$\sqrt{4}$ = 41.5 & 9.3 & 20.76*$\sqrt{8}$ = 58.7 & 17.5 \\ \cline{3-7}
      & & 1.0 &  20.76*$\sqrt{4}$ = 41.5 & 57.7 & 20.76*$\sqrt{8}$ = 58.7 & 92.0 \\ \hline     
    \end{tabular}
    \caption{Fit results of the study with scaled $\delta$ and scaled fracture energy G}
    \label{tab:fitsGScaled}
\end{table}

Considering both scaling schemes, it is evident that regardless of the critical stretch scaling scheme, $a_0$ is significantly influenced by the length scale of the model, while $K_c$ depends on the length scale insofar as the fracture energy is changed. 

\begin{figure}[tbh]
    \centering
    \includegraphics[width=\textwidth]{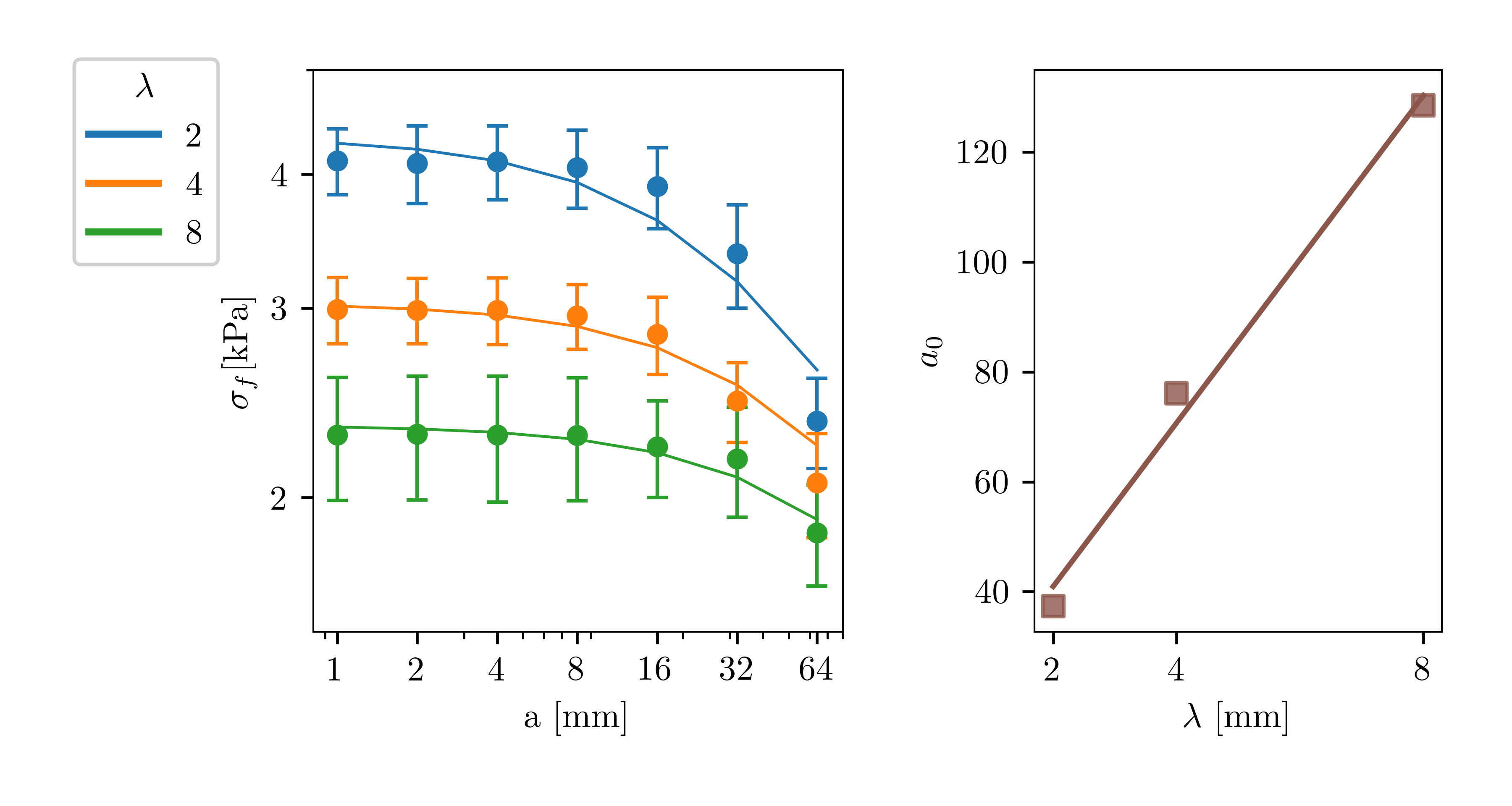}
    \caption{Left: failure stress vs a for simulations with a fixed horizon of 4 mm and varying correlation lengths, curve fitted using model cH2; right: dependence of $a0$ on correlation length $\lambda$}
    \label{fig:fitsCorrlength}
\end{figure}

So far we considered the horizon and correlation length to be always equal. Next, we examine situations where the two length scales are varied independently. To begin with, we consider a fixed horizon $\delta$ and systematically change the correlation length of the random field $\lambda$. In the simulations, we keep the system size, disorder parameter, and horizon fixed at $L = 400$ mm, $\alpha = 1.0$, $\delta$ = 4 mm, while we vary the correlation length in the range $\xi \in$ [2mm, 4mm, 8mm]. Failure stresses from these simulations are shown in \autoref{fig:fitsCorrlength}, left. 
For analyzing the data we test the following hypotheses:  
(cH1) The data are best described by a $K_c$ and $a_0$ that depend both on $\xi$; (cH2) the data are best described by a common $K_c$ while $a_0$ depends on $\lambda$; (cH3) $K_c$ depends on $\lambda$ but $a_0$ is independent on $\lambda$.  
\begin{table}[h]
    \centering
    \begin{tabular}{|*{8}{c|}}
    \hline
    \multirow{3}{3em}{Model} & \multirow{3}{3em}{AIC} & \multicolumn{6}{|c|}{$\lambda$} \\ \cline{3-8}
      &  & \multicolumn{2}{|c}{2 mm} & \multicolumn{2}{|c|}{4 mm} & \multicolumn{2}{c|}{8 mm} \\ \cline{3-8}
      & & $K_c$ & $a_0$ & $K_c$ & $a_0$ & $K_c$ & $a_0$ \\ \hline 
      cH1 & -498.4 & 48.5 & 40.3 & 41.5 & 57.8 & 47.9 & 134.2 \\ \hline
      cH2 & -499.1 & 46.9 & 37.4 & 46.9 & 76.1 & 46.9 & 128.46 \\ \hline    
      cH3 & -481.6 & 54.3 & 52.8 & 40.0 & 52.8 & 31.8 & 52.8 \\ \hline
    \end{tabular}
    \caption{Fit results of the study with fixed $\delta$ and varying $\lambda$} 
    \label{tab:fitsXi}
\end{table}
Comparing the AIC of the three models, it is clear that cH2 best describes the data. The resulting dependency of $a_0$ on $\lambda$ is approximately linear, as illustrated in \autoref{fig:fitsCorrlength}, right.

\begin{figure}[tbh]
    \centering
    \includegraphics[width=\textwidth]{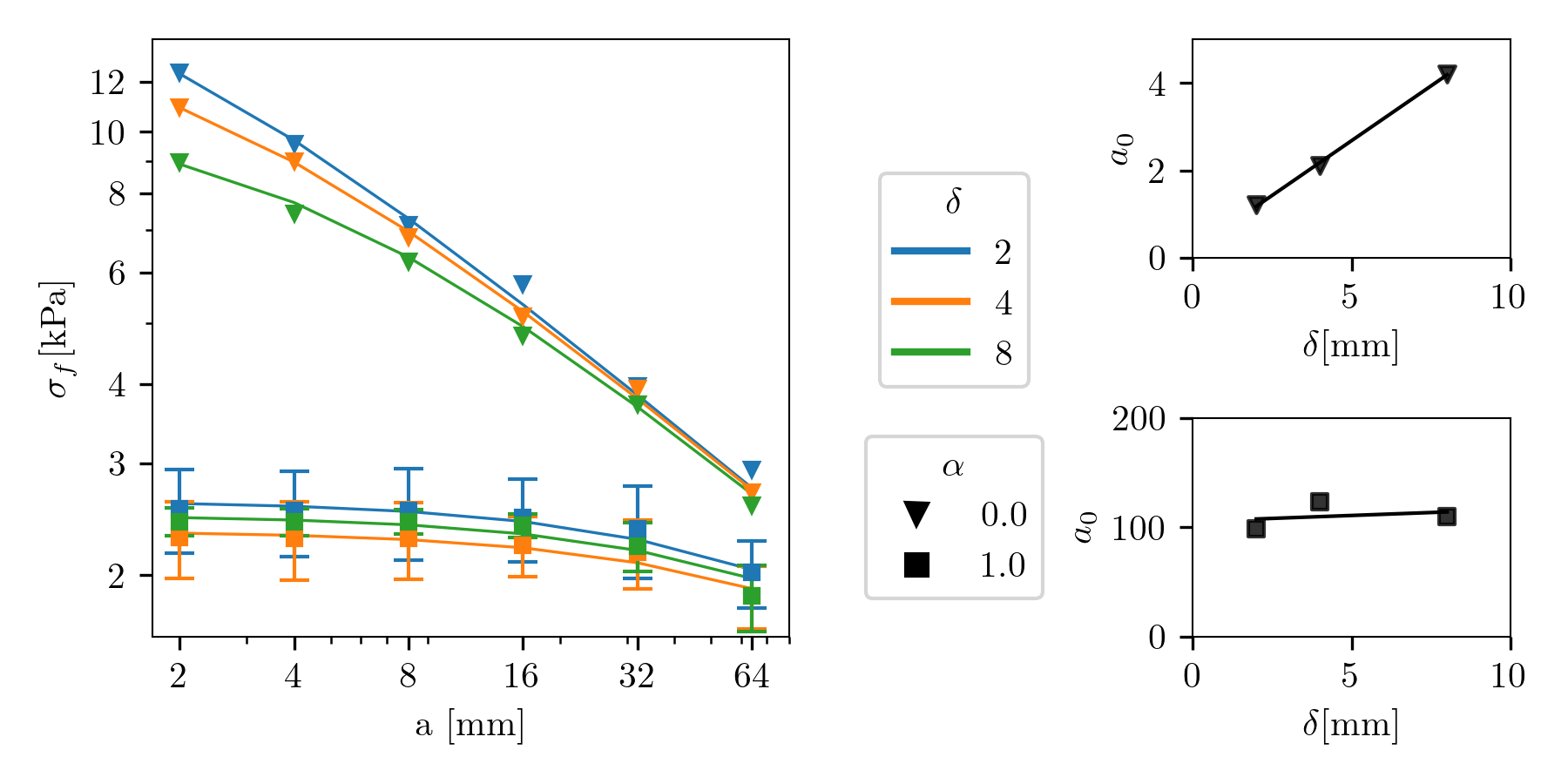}
    \caption{Left: failure stress vs notch length $a$ for simulations with  fixed correlation length $\lambda = 8$mm and varying horizons, fitted using parameters obtained for dH2; right: dependence of $a_0$ on horizon for no disorder (top, $\alpha = 0$) and high disorder (bottom, $\alpha = 1$), fit parameters obtained from dH2.}
    \label{fig:fitsDeltaNotXi}
\end{figure}

Finally, we consider situations where the fluctuation correlation length $\lambda$ is kept fixed but the horizon $\delta$ is changed. When changing $\delta$, we also scale $s_c$ in order to keep the fracture energy G fixed, so that according to the previous studies, $K_c$ is expected to remain constant. Simulations were performed for $L = 400$ mm, $\alpha \in$[0.0, 1.0], $\xi$ = 8 mm, $\delta \in $[2mm, 4mm, 8mm], with $s_c$ = 0.0099, 0.007 and 0.0049 respectively. 
Results are compiled in \autoref{fig:fitsDeltaNotXi}.
To analyse the data, we consider two hypotheses, namely (dH1) that both $K_{\rm c}$ and $a_0$ depend on $\delta$, and (dH2) that only $a_0$ depends on $\delta$. As shown in \autoref{tab:fitsDelta}, (dH2) is slightly preferred according to AIC, and is also in line with our previous findings that point to an approximately constant value of $K_{\rm c}$. 

\begin{table}[h]
    \centering
    \begin{tabular}{|*{9}{c|}}
    \hline
    \multirow{3}{3em}{Model} & \multirow{3}{3em}{AIC} & \multirow{3}{3em}{$\alpha$} & \multicolumn{6}{|c|}{$\delta$} \\ \cline{4-9}
      &  & & \multicolumn{2}{|c}{2 mm} & \multicolumn{2}{|c|}{4 mm} & \multicolumn{2}{|c|}{8 mm}\\ \cline{4-9}
      & & & $K_c$ & $a_0$ & $K_c$ & $a_0$  & $K_c$ & $a_0$\\ \hline 
      \multirow{2}{3em}{dH1} & -60.3 & 0.0 & 40.2 & 1.4 & 38.9 & 2.0 & 38.0 & 3.7 \\ \cline{2-9}
      & -455.3& 1.0 & 49.1 & 113.6 & 46.9  & 127.5 & 42.5 & 91.2 \\ \hline 
      \multirow{2}{3em}{dH2} & -60.7 & 0.0 & 39.3 & 1.2 & 39.3 & 2.1 & 39.3 & 4.2\\ \cline{2-9}
      &-458.4 & 1.0 & 46.1 & 98.7 & 46.1  & 122.9 & 46.1 & 109.6 \\ \hline  
    \end{tabular}
    \caption{Fit results of the study with fixed $\lambda$=8.0 mm and varying $\delta$}
    \label{tab:fitsDelta}
\end{table}
In absence of disorder, $a_0$ increases systematically with $\delta$ and the fitted values point to an approximately linear proportionality. At high disorder, the data for different values of $\delta$ overlap within the statistical margins. In this case the fitted values of the process zone size $a_0$ as shown in \autoref{tab:fitsDelta} do not  exhibit a systematic dependency on $\delta$ but scatter around the value obtained from dH1 in a non-systematic manner, indicating that one is actually 'fitting the noise'. 

\begin{figure}[tbh]
    \centering
    \includegraphics[width=\textwidth]{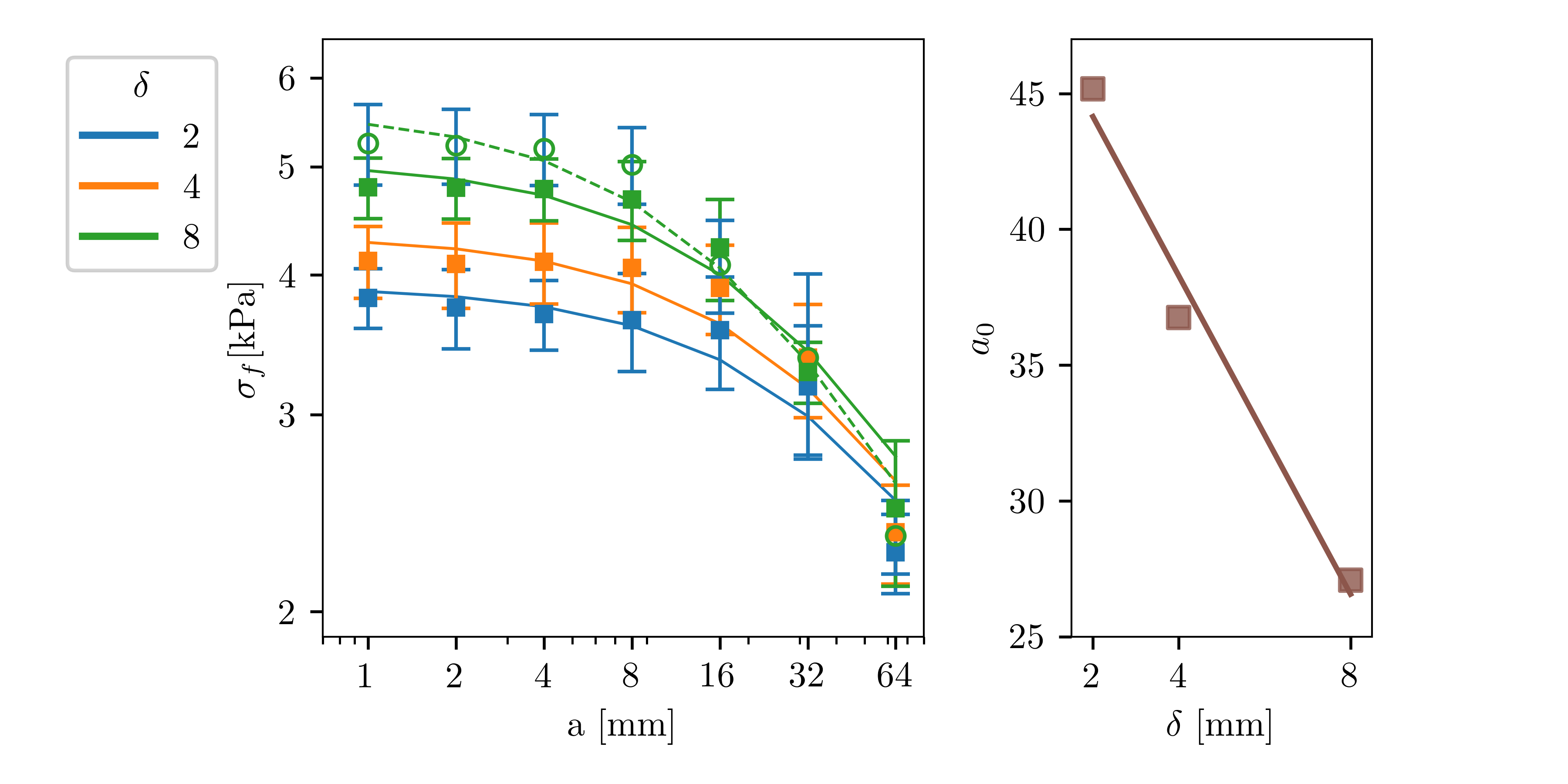}
    \caption{Left: failure stress vs notch length $a$ for simulations with  fixed correlation length $\lambda = 2$mm and varying horizons, fitted using parameters obtained for eH2; right: dependence of $a_0$ on horizon for high disorder ($\alpha = 1, CV = 1.31$); a curve for $\alpha = 0.5, CV=0.53$ evaluated with $\lambda = \delta = 8$mm is shown for comparison (open circles).}
    \label{fig:fitsDeltaNotXismall}
\end{figure}

All data shown in \autoref{fig:fitsDeltaNotXi} for the case of high disorder are obtained for situations where the fluctuation correlation length exceeds the peridynamic horizon. But what happens in the opposite limit where the horizon exceeds the fluctuation correlation length? To study this problem, we performed a range of simulations in the limit of high disorder with $\lambda = 2$mm, $\delta \in [$2mm, 4mm, 8mm]. The results shown in \autoref{fig:fitsDeltaNotXismall} show the apparently paradoxical result that in this case, the process zone size $a_0$ {\em decreases} with increasing horizon size. This behavior is observed independently of whether both $a_0$ and $K_{\rm c}$ are assumed to depend on $\delta$ (eH1) or only $a_0$ is assumed to depend on $\delta$ (eH2). We note that, as in most other cases, assuming the fracture toughness to be independent on internal length scales yields a better representation of the data (see AIC values and fit parameters given in \autoref{tab:fitsDeltaXismall}). 

To understand the reasons for the decrease of $a_0$ with increasing $\delta$, we observe that, for $\delta > \lambda$, the non-local evaluation of interactions leads to a situation where the fluctuating local modulus is averaged over the horizon $\delta$, which acts as effective correlation length of the averaged random field. In other words, a high degree of disorder with a correlation length $\lambda \ll \delta$ is approximately equivalent to a reduced degree of disorder with a correlation length $\lambda \approx \delta$: increasing $\delta$ for $\lambda \ll \delta$ is equivalent to decreasing the degree of disorder. This, in turn, implies a lower value of $a_0$. To illustrate the point, we have added, in \autoref{fig:fitsDeltaNotXismall}, data obtained for $\delta = \lambda = 8$mm obtained with a reduced value of $\alpha=0.5$ (open circles in \autoref{fig:fitsDeltaNotXismall}).

\begin{table}[h]
    \centering
    \begin{tabular}{|*{9}{c|}}
    \hline
    \multirow{3}{3em}{Model} & \multirow{3}{3em}{AIC} & \multirow{3}{3em}{$\alpha$} & \multicolumn{6}{|c|}{$\delta$} \\ \cline{4-9}
      &  & & \multicolumn{2}{|c}{2 mm} & \multicolumn{2}{|c|}{4 mm} & \multicolumn{2}{|c|}{8 mm}\\ \cline{4-9}
      & & & $K_c$ & $a_0$ & $K_c$ & $a_0$  & $K_c$ & $a_0$\\ \hline 
      eH1 & -413.1 & 1.0 & 46.9 & 46.0 & 48.5 & 40.3 & 45.3 & 25.3 \\ \hline
      eH2 & -416.0 & 1.0 & 46.6 & 45.4 & 46.6 & 36.8 & 46.6 & 27.1\\ \hline
    \end{tabular}
    \caption{Fit results of the study with fixed $\lambda$=2.0 mm and varying $\delta$}
    \label{tab:fitsDeltaXismall}
\end{table}

To summarize, our study of the effect of internal length scales on the fracture behavior of disordered samples shows a complex behavior. In samples without disorder, the process zone size is proportional to the peridynamic horizon, and $\delta$ and $a_0$ are of the same order of magnitude. In disordered samples, the process zone size increases with increasing disorder (increasing scatter of local strength) and may become much larger than $\delta$ in heavily disordered samples. In the regime of moderate and large disorder, $a_0$ increases with increasing fluctuation correlation length $\lambda$, whereas in this regime it shows no statistically significant dependency on $\delta$. An interesting observation is that, in highly disordered regime, the process zone size increases with system size, i.e., it does not simply represent a material characteristic.

Regarding fracture toughness, we find that it does not depend significantly on either the degree of disorder or on the internal length scales  $\delta$ and $\xi$, unless the simulations are carried out such as to keep the critical strain $s_{\rm c}$ independent on $\delta$. In this case one observes a proportionality $K_{\rm c} \propto \sqrt{\delta}$ as expected from fracture mechanics and from peridynamic theory of disorder-free materials.

\section{Discussion and Conclusions}

When disorder is introduced into a peridynamic model by considering the material parameters as statistically homogeneous random fields ('intermediate homogenization'), then this procedure necessarily introduces additional length scales in terms of the correlation length(s) of these fields. Together with the internal length scale given by the peridynamic horizon, such intermediately homogenized models inherently exhibit a multi-scale nature. As such they are suitable for modelling disordered microstructures where the range of elementary interactions may in general differ from the correlation length of the microstructure. 

An example is provided by disordered alloys such as high-entropy alloys, where the disordered arrangement of chemically distinct atoms introduces composition and stress fluctuations on multiple scales within a crystal lattice \citep{wu2025dislocations}. While the equivalent of the peridynamic horizon is provided by the range of atomic interactions, the correlation length of composition fluctuations, internal stresses or local (chemical or structural) ordering in such materials may be much larger and can moreover be tweaked by processing, e.g. through thermal treatment \citep{liang2024controllable}. A second example are anmorphous metals (metallic glasses) where, depending on processing, local crystalline inclusions may modulate the mechanical behavior on tunable scales \citep{eckert2007mechanical}. Moving to much larger scales, the elementary spatial scale of interactions in architected metamaterials is defined by the unit building blocks, such as elementary cells or beams, whereas fluctuations in form of random modulations of the metamaterial geometry can be designed on arbitrary scales \cite{zaiser2023disordered}. 

Our investigation has demonstrated that, in quasi-brittle failure of such multi-scale materials, the interplay between the length scale of interactions and the correlation length of fluctuations, in conjunction with the interplay between the effects of random stress and strength fluctuations and crack-tip stress concentrations, may give rise to complex behavior. Generally, we find that the dependency of strength on crack length is well described by the McClintock-Irwin relation, and that the fracture toughness does not strongly depend on either the internal length scales or the degree of microstructural disorder. However, the process zone size, which controls the nature of the failure process (crack propagation vs distributed damage nucleation and accumulation), exhibits a complex dependency on both the degree of disorder, and on external and internal length scales. 

The situation is simple in absence of disorder. In this case, the process zone size is approximately proportional to the horizon, $a_0 \approx \kappa \delta$ with a proportionality coefficient that is, for a constant micro-modulus, around $\kappa = 0.5$ (\autoref{fig:fitsDeltaNotXi}, top curves, and parameters in \autoref{tab:fitsDelta}). In disordered samples, the size of the process zone increases significantly with an increasing degree of disorder, and the behavior becomes more complex. If $\delta = \lambda$ (model with a single length scale), the process zone size $a_0$ is approximately proportional to but much larger than $\delta = \lambda$, irrespective of the scaling rule adopted (constant $G$ or constant $s_{\rm c}$, see \autoref{fig:fitsLengthScale} and $a_0$ values given in \autoref{tab:fitsGFixed} and \autoref{tab:fitsGScaled}). If $\delta \gg \lambda$, then the situation is similar to $\delta = \lambda$ but with a degree of disorder that decreases with increasing $\delta/\lambda$ ratio (see \autoref{fig:fitsDeltaNotXismall}). Finally, in disordered samples where $\lambda > \delta$, the process zone size is approximately independent of $\delta$ (\autoref{fig:fitsDeltaNotXi}, bottom curves) but increases approximately linearly with $\lambda$ (\autoref{fig:fitsCorrlength} and \autoref{tab:fitsXi}). 

Beyond internal length scales, we observe a dependency of the damage patterns on the size of the simulated samples in form of a  statistically significant, approximately logarithmic increase of the process zone size with increasing system size $L$ (\autoref{fig:fitsL} and \autoref{tab:fitsAlphaL}). This observation holds even when $L$ is significantly larger than all internal length scales. This observation is matched by a logarithmic size effect in samples without pre-existing cracks (\autoref{fig:sizeEffect}). In fact, the McClintock-Irwin relation predicts the failure strength of intact samples to be inversely proportional to $\sqrt{a_0}$, hence the logarithmic size effect demonstrated in \autoref{fig:sizeEffect} necessarily implies a logarithmic size dependency of the process zone. Looking at the spatial distribution of damage prior to failure, we observe that in highly disordered samples damage accumulation may spread widely across the sample (\autoref{fig:damagepattern} and \autoref{fig:a0vs_alpha_L},right). This may lead to features typical of highly disordered materials such as an increase in the non-essential work of failure and a statistical size effect related to nucleation of cracks at extended zones of low strength acting as 'weak spots'. In extremely disordered samples, \citet{shekhawat2013damage} reported a transition to percolation-like behavior. We do not observe such a transition but it remains to be investigated whether it might occur at even higher degrees of disorder than investigated here. 

Another question for further investigation is the development of systematic relations describing how the mesoscopic statistical features (correlation length and probability distribution of local properties) emerge from coarse-graining of the actual, fully resolved microstructure of a porous or composite material. While the present investigation focuses on the structure and properties of intermediately homogenized models, their systematic derivation through statistical coarse graining remains an important task for future investigation. 

\section*{Data availability}
Datasets generated and analyzed during this study are published and available on Zenodo, 10.5281/zenodo.17898813.

\section*{Conflict of interest}
The authors declare that they have no conflict of interest.

\section*{Funding}
This research was funded by the Deutsche Forschungsgemeinschaft (DFG, German Research Foundation) - 377472739/GRK 2423/2-2023 and by grant Mo 3049/3-3. The authors are very grateful for this support.

\bibliographystyle{spbasic}      
\bibliography{references_disorder}   

\end{document}